\documentclass[11pt,a4paper]{article}

\pdfoutput=1
\def\bq{\begin{equation}}
\def\ee{\end{equation}}
\def\bea{\begin{eqnarray}}
\def\eea{\end{eqnarray}}
\def\equ#1{(\ref{#1})}
\usepackage{epsfig,jheppub}
\def\FIGURE#1{\begin{figure}[ht]#1\end{figure}}

\title{A Transfer Matrix Method for Resonances in Randall-Sundrum Models}

\author[a]{R. R. Landim,}
\author[b]{G. Alencar,}
\author[b]{M. O. Tahim,}
\author[a]{and R. N. Costa Filho}
\affiliation[a]{Departamento de F\'{\i}sica, Universidade Federal do Cear\'{a},
Caixa Postal 6030, Campus do Pici, 60455-760, Fortaleza, Cear\'{a}, Brazil.\\}
\affiliation[b]{Universidade Estadual do Cear\'a, Faculdade de Educa\c c\~ao, Ci\^encias e Letras do Sert\~ao Central-R. 
Epitácio Pessoa, 2554, 63.900-000  Quixad\'{a}, Cear\'{a},  Brazil.
}

\emailAdd{geovamaciel@gmail.com}
\emailAdd{renan@fisica.ufc.br}
\emailAdd{mktahim@yahoo.com.br}
\emailAdd{rai@fisica.ufc.br}

\abstract{
	In this paper we discuss in detail a numerical method to study resonances in membranes generated by domain walls in Randall-Sundrum-like scenarios. It is based on similar works to
 understand the quantum mechanics of electrons subject to the potential barriers that exist in heterostructures in semiconductors. This method was used recently to study resonances of a 
three form field and lately generalized to arbitrary forms. We apply it to a lot of important models, namely those that contain the Gauge, Gravity and Spinor fields. In many cases we find a 
rich structure of resonances which depends on the parameters involved.}

\dedicated{This paper is dedicated to the memory of my wife  Isabel Mara (R. R. Landim)}

\begin{document}
\maketitle
\section{Introduction}

		It is a well known result that when electrons move inside semiconductors, they experiment the potential barriers due to the heterostructures of the material. With 
basics quantum mechanics, by solving the Schroedinger equation we can compute transmission and reflection coefficients, making possible the understanding of the resonance structures of 
each case of potentials studied. This is one of the aspects of theoretical condensed matter extensively studied in the 90's \cite{kim:1988,renan:1993,renan:1994}. Due to the improvement of 
numerical computational methods,  a few more difficult problems without exact solutions were attacked. The source of those difficulties is the complexity of the several kinds of potentials 
that should be treated in the Schroedinger equation. In particular, a numerical method was developed in order to solve the problem of potential barriers for any case \cite{ando:1987}. In order 
to obtain these results numerically, they made use of a common tool in condensed matter systems: The transfer matrix. In a few words, the program computes transmission coefficients for the 
given potential profile. 

	The fact is that, in the last years, the models in physics of extra dimensions brought back to the arena the analysis of the Schroedinger equation with a potential generated by 
thin membranes. In these models, the interaction between the several fields and the membrane is an important aspect to be understood. For example, localization of fields and the structure 
of resonances can tell how the the universe inside the membrane behaves. In that sense, particles moving through the extra dimensions resembles the wave packets of electrons hitting the 
potential barriers inside semiconductors. The core idea of extra dimensional models is to consider the four-dimensional universe as a hypersurface embedded in a multidimensional manifold. 
In particular, the standard model presents interesting topics to study such as the hierarchy problem, and the cosmological constant problem that can be treated by the above-mentioned scenarios.
For example, the Randall-Sundrum (RS) model \cite{Randall:1999vf,Randall:1999ee} provides a possible solution to the hierarchy problem and show how gravity is trapped to a membrane.

	For compact extra dimensions, as in Kaluza-Klein models, a massless field in five dimensions gives origin to a massive one in four dimensions. The spectrum of masses is discrete and they are determined by the radius of the extra dimension. In RS models the extra dimension is infinity. In these models, generally the mass spectrum is determined by a Schroedinger like 
equation and the spectrum is not guaranteed to be discrete. In these models it is important to verify if electromagnetism and gravity are reproduced in first approximation inside the brane. 
This is true if the zero mode, or massless mode, is localized in the membrane. In these scenarios, some facts about localization of fields are known: The scalar field ($0-$form) and the 
gravity field are localizable, but the vector gauge fields ($1-$form) is not. The reason why this happens is that, in four dimensions, the vector field theory is conformal and all information 
coming from warp factors drops out, necessarily rendering a non-normalizable four dimensional effective action. Beyond this the left hand of the fermion field is also localizable. 

	It is important to note that the presence of one more extra dimension ($D=5$) provides the existence of many antisymmetric fields, namely the two, three, four and five forms. However, 
the only relevant ones for our membrane are the two and the three form. This is due to the fact that when the number of dimensions increase, also increases the gauge freedom. This can be used 
to cancel the dynamics of the field in the visible brane. From the physical viewpoint, they are of great interest because they may have the status of fields describing particles other than the 
usual ones. As an example we can cite the space-time torsion \cite{Mukhopadhyaya:2004cc} and the axion field \cite{Arvanitaki:2009fg,Svrcek:2006yi} that have separated descriptions by the 
two-form. Besides this, String Theory shows the naturalness of higher rank tensor fields in its spectrum \cite{Polchinski:1998rq,Polchinski:1998rr}. Other applications of these kind
of fields have been made showing its relation with the AdS/CFT conjecture \cite{Germani:2004jf}.

	For these fields it is also a known fact that their zero mode are not localized. The mass spectrum of the two and three form have been studied, for example, 
in Refs. \cite{Mukhopadhyaya:2004cc} and \cite{Mukhopadhyaya:2007jn}. Posteriorly, the coupling between the two and three forms with the dilaton was studied, in different contexts, 
in \cite{DeRisi:2007dn,Mukhopadhyaya:2009gp,Alencar:2010mi}. The study of this kind of coupling, inspired in string theory, is important in order to produce a process that, in principle, 
could be seen in LHC. This is a Drell-Yang process in which a pair quark-antiquark can give rise to a three (two)-form field, mediated by a dilaton. 

	In this approach, we make use of soliton-like solutions: these are studied with increasing interest in physics, not only in Condensed Matter, as in Particle Physics and Cosmology. 
In brane models, they are used as mechanism of field localization, avoiding the appearance of the troublesome infinities. Several kinds of defects in brane scenarios are considered in the
literature \cite{Gremm:1999pj,yves:a,yves:b,yves:c}. As an example, in a recent paper, a model is considered  for coupling fermions to brane and/or antibrane modeled by a kink antikink 
system \cite{yves:d}. The localization of fields in a framework that consider the brane as a kink have been studied for example in 
\cite{Bazeia:2008zx,Bazeia:2007nd,Bazeia:2004yw,Bazeia:2003aw,Christiansen:2010aj}.

	A proposal to solve the gauge field localization problem was given by Kehagias and Tamvakis \cite{Kehagias:2000au}. They have shown that the coupling between the dilaton and the 
vector gauge field produces the desired localization. In this work they also consider the localization of gravity in the background with and without the dilaton. In analogy with their work, 
the present authors have also considered the issue of localization and resonances of a three-form field in \cite{Alencar:2010hs} and in two separate papers,  $q-$form fields in a scenario of 
a $p-$brane with co-dimension one and two \cite{Landim:2010pq,Alencar:2010vk}. As has been pointed out in \cite{Landim:2010pq}, the coupling here is needed for a different reason. In that work, 
it was shown that the some form fields without a dilaton do not have its massless modes localized. When we introduce a dilaton, this mode can be localized. 

	Therefore we are left with two phenomenological setups: with and without the dilaton field. In the case of form fields, only the zero form do not need of the dilaton coupling. When 
we consider the other fields, namely the gravity and fermion fields, the dilaton is not needed for localization. In the recent literature, generally the dilaton is considered in the form 
field case and not in the gravity and fermion cases.  Therefore a more complete analyses should consider all the fields in both setups. 

	In all the models cited above a common feature is that the massive modes are not normalizable and therefore we do not have a well defined effective action. Despite of this 
there is the important possibility of appearance of resonances. As said before the mass spectrum is determined by a Schroedinger like equation. The potential term in this equation will 
generally have a volcano shape and therefore we can ask about the possibility of massive modes, besides living in the extra dimension, have a peak of probability to be found at the location 
of the membrane. This analysis have been done extensively in the literature \cite{Bazeia:2005hu,Bazeia:2002xg,Liu:2009ve,Zhao:2009ja,Liang:2009zzf,Zhao:2010mk,Zhao:2011hg,Li:2010dy,Castro:2010au,Correa:2010zg,
Castro:2010uj,Chumbes:2010xg,Castro:2011pp,Almeida:2009jc,Cruz:2010zz,Cruz:2009ne} . For example, in \cite{Cruz:2010zz}, the authors consider a the vector field and found that, after 
normalization of the the wave function in a truncated region, there is a resonance very close to $m=0$, and this indicates that the lighter is the mass, the bigger is the probability 
of interaction with the membrane. 

	From our viewpoint the wave function has an oscillatory part and cannot be normalized. In order to study this, the authors define a relative probability, but for us it is not 
clear how it solves the problem. Another very important point is the truncation of the integration region. This is equivalent to have a potential that do not falls to zero at infinity, 
therefore it is natural to have a peak of resonance very close to $m=0$. This could lead to the wrong interpretation that very light modes can interact with the membrane. From our analysis, 
it is
not true that the lighter are the massive modes, the bigger is the probability of finding them inside the membrane. In fact, using our method it has been show \cite{Alencar:2010hs,Landim:2010pq} 
that heavy modes also can interact with the membrane differently from what is discussed in \cite{Cruz:2009ne}. In fact we have found a very rich structure of resonances.

	In order to analyse resonances, we must compute transmission coefficients ($T$). This gives a clearer and cleaner physical interpretation about what happens to a free wave that 
interact with the membrane. The idea of the existence of a resonant mode is that for a given mass the transmitted and reflected oscillatory modes are in phase inside the membrane, i. e., the transmission coefficient has a peak at this mass value. That means the amplitude of the wave-function has a maximum value at $z=0$ and the probability to find this KK mode inside the membrane is higher.

	Therefore the main subject of this article is to give in detail the numerical method used recently to study aspects of the form fields in a scenario of extra dimensions. As said 
before, it is lacking in literature the study of resonances for form fields without the dilaton and for gravity and fermion fields with the dilaton. Therefore as an application of the above 
method we compute the transmission coefficients of Gravity, Fermion and form fields in both setups: with and without the dilaton coupling.  The paper is organized as follows. Section two 
is devoted to discuss a solution of Einstein`s equation with a source given by a kink with and without the presence of the dilaton. In section two we give the general prescription to reach 
the Schroedinger equation used to analyses the problem. We also present in detail the numerical steps to compute the transmission coefficients with the transfer matrix.  In Section four we 
analyses resonances of the gravity field. In section five, six and seven, we present the resonance structure 
for $0$,$1$ and $2$ forms respectively. In section eight we present the same study for the fermion fields. At the end, we discuss the conclusions and 
perspectives.

\section{The Kink as a Membrane}

	We start our analysis by studying the space-time background. 
It is well known that vector gauge fields in these kind of
scenarios are not localizable: in four dimensions the gauge vector field
theory is conformal and all information coming from warp factors drops out
necessarily rendering a non-normalizable four dimensional effective action.
However, in the work of Kehagias and Tamvakis \cite{Kehagias:2000au}, it is shown
that the coupling between the dilaton and the vector gauge field produces
localization of the later. On the other hand scalar and fermion fields do 
not need this coupling in order to produce localized zero modes. Therefore we must consider in this section both backgrounds: with and without the dilaton coupling. 

\subsection{The Background with the Dilaton Coupling}

	In this section we obtain a solution of the equations of motion for the gravitational 
field in the background of the dilaton and the membrane. For such, we introduce the
following action, similar of the one used in \cite{Kehagias:2000au}:
\begin{equation}
S=\int d^{5}x \sqrt{-G}[2M^{3}R-\frac{1}{2}(\partial\phi)^{2}-\frac{1}{2}%
(\partial\pi)^{2}-V(\phi,\pi)],
\end{equation}
where $G$ is the metric determinant and $R$ is the Ricci scalar. Note that we are
working with a model containing two real scalar fields. The field $\phi$ plays the role of
membrane generator of the model while the field $\pi$ represents the dilaton. The
potential function depends on both scalar fields. It is assumed the following ansatz for
the space-time metric:
\begin{equation}
ds^{2}=e^{2A(y)}\eta_{\mu\nu}dx^{\mu}dx^{\nu}+e^{2B(y)}dy^{2},
\end{equation}
where $\eta_{\mu\nu}=diag(-1,1,1,1)$ is the metric of the $p$-brane, $y$ is the
co-dimension coordinate. As usual, capital Latin indexes represent the coordinates in the bulk and Greek indexes, those on the brane.
The equations of motion are given by
\begin{equation}
\frac{1}{2}(\phi^{\prime})^{2}+\frac{1}{2}(\pi^{\prime})^{2}-e^{2B(y)}V(%
\phi,\pi)=24M^{3}(A^{\prime})^{2},
\end{equation}
\begin{equation}
\frac{1}{2}(\phi^{\prime})^{2}+\frac{1}{2}(\pi^{\prime})^{2}+e^{2B(y)}V(%
\phi,\pi)=-12M^{3}A^{\prime\prime}-24M^{3}(A^{\prime})^{2}+12M^{3}A^{%
\prime}B^{\prime},
\end{equation}
\begin{equation}
\phi^{\prime\prime}+(4A^{\prime}-B^{\prime})\phi^{\prime}=\partial_{\phi}V,
\end{equation}
and
\begin{equation}
\pi^{\prime\prime}+(4A^{\prime}-B^{\prime})\pi^{\prime}=\partial_{\pi}V.
\end{equation}
In order to solve this system, we use the so-called super-potential function $
W(\phi)$, defined by $\phi^{\prime}=\frac{\partial W}{\partial\phi}$,
following the approach of Kehagias and Tamvakis \cite{Kehagias:2000au}. The
particular solution regarded follows from choosing the potential $V(\phi,\pi)$ and super-potential $W(\phi)$ as
\begin{equation}
V=\exp{(\frac{\pi}{\sqrt{12M^{3}}})}\{\frac{1}{2}(\frac{\partial W}{%
\partial\phi})^{2}-\frac{5}{32M^{3}}W(\phi)^{2}\},
\end{equation}
and
\begin{equation}
W(\phi)=va\phi\left(1-\frac{\phi^2}{3v^2}\right),
\end{equation}

where $a$ and $v$ are parameters that adjust the dimensionality. As pointed out in \cite{Kehagias:2000au}, this potential give us the desired soliton-like solution. In this way it is easy to obtain first order differential equations whose solutions are solutions of the
equations of motion above, namely
\begin{equation}
\pi=-\sqrt{3M^{3}}A,
\end{equation}
\begin{equation}
B=\frac{A}{4}=-\frac{\pi}{4\sqrt{3M^{3}}},
\end{equation}
and
\begin{equation}
A^{\prime}=-\frac{W}{12M^{3}}.
\end{equation}
The solutions for these new set of equations are the following:
\begin{equation}
\phi(y)=v\tanh(ay),  \label{dilat}
\end{equation}
\begin{equation}
A(y)=-\frac{v^2}{72M^3}\left(4\ln\cosh(ay) +\tanh^2(ay)\right) \label{dilat2}
\end{equation}

All of these results are going to be used for analyses resonances of several fields with $v^2/72M^3=1$.

\subsection{The Background without the Dilaton Coupling}

Here we look for the background solution without the dilaton coupling. We could repeat all the steps of the last section to arrive at the final solution. We can also take a short way and from our previous solution we obtain the desired result. By just setting $B=\pi=0$ we can get the final answer. Therefore we get for the metric
\begin{equation}
ds^{2}=e^{2A(y)}\eta_{\mu\nu}dx^{\mu}dx^{\nu}+dy^{2},
\end{equation}
and for the resulting potential
\begin{equation}
V=\{\frac{1}{2}(\frac{\partial W}{%
\partial\phi})^{2}-\frac{5}{32M^{3}}W(\phi)^{2}\}.
\end{equation}
This solution agrees with the one obtained previously. We must note that the solution for $A$ is left unchanged and therefore the contribution of the dilaton is to change the equations of motion. Therefore we end here the analysis of this background and in the next section we present the general set to be used throughout this paper in order to look for the spectrum of resonances of fields. 

\section{The Numerical Method}

In this section we give the general prescription to be followed in all the cases to be studied in this work. First we must show how to reach the equation that drives the massive modes and argue that it is a Schroedinger-like equation. Next we show in detail the transfer matrix method to compute transmission coefficients of an arbitrary potential profile. 

\subsection{The Schroedinger-like Equation}
The first step is to find the equation of motion for the field in five dimensions, which generally takes the form
\begin{equation}
\hat{O}\Phi(x,y)=0, 
\end{equation} 
where $\hat{O}$ is a differential operator em five dimensions. The next step is to separate the differential operator in the brane ($\hat{O}_{4d}$) and extra dimension($\hat{O}_y$):
\begin{eqnarray}
\hat{O}=\hat{O}_{4d}+\hat{O}_y.
\end{eqnarray} 
After this we perform a separation of variables in the field $\Phi(x,y)=\psi(y)\phi(x)$. It is important to note that when doing this all the information about the polarizations will be in $\phi$ and $\psi$ will be a scalar function. After these considerations we get
$$
\frac{\hat{O}_{4d}\phi(x)}{\phi(x)}=-\frac{\hat{O}_y\psi(y)}{\psi(y)}  
$$
and as the lhs depends only on $x$ and the rhs on $y$ they both must be equal to a constant that we call $m^2$. The motivations for this constant is obvious after we write the equations of motions as
$$
\hat{O}_{4d}\phi(x)=-m^2\phi(x)
$$
and
$$
\hat{O}_y\psi(y)=m^2\psi(y).
$$
The solutions of the second equation above therefore give us the allowed masses in the visible brane. This equation in all cases will have the form
\begin{equation}
 \left(-\frac{d^2}{dy^2}+P'(y)\frac{d}{dy}+V(y)\right)\psi(y)=m^2Q(y)\psi(y).\label{pq}
\end{equation}
The best way to study this equation is to transform it in a Schroedinger type equation
\begin{equation}
 \left(-\frac{d^2}{dz^2}+{U}(z)\right)\overline{\psi}(z)=m^2\overline{\psi}(z),\label{schlike}
\end{equation}
through the transformations
\begin{equation}
 \frac{dz}{dy}=f(y), \quad \psi(y)=\Omega(y)\overline{\psi}(z),
\end{equation}
with
\begin{equation}
 f(y)=\sqrt{Q(y)}, \quad \Omega(y)=\exp(P(y)/2)Q(y)^{-1/4},
\end{equation}
and
\begin{equation}
 {U}(z)=V(y)/f^2+\left(P'(y)\Omega'(y)-\Omega''(y)\right)/\Omega f^2 \label{potential}
\end{equation}
where the prime is a derivative with respect to $y$. In order to obtain a meaningful effective action we also need the integration in the extra dimension be finite. After the above transformation of coordinates, the interesting thing is that this finiteness is translated in the condition
$$
\int \overline{\psi}^2 dz=finite
$$
which is like the square integrable condition of quantum mechanics. Therefore our problem can be addressed as if we were dealing with an one dimension quantum mechanical problem with potential given by \equ{potential}. The study of the localization of massless modes ($m^2=0$) is very simple and it has been widely studied in the literature. An obvious solution to the massless case is the constant mode $\psi=\psi_0$. With this the condition for localization reduces to
$$
\int \Omega dy=finite.
$$
The study of the massive modes must be done more carefully. This happens because when we go to specific cases we generally get that the potential has the asymptotic behaviour
$$
\lim _{z \rightarrow \pm\infty}U(z)=0.
$$
Therefore we do not have a discrete spectrum for $m>0$. Beyond this, it is a known fact from quantum mechanics that any solution for positive $m$ must posses a oscillatory contribution, and therefore the wave function is not normalizable. The conclusion is that the only possible localized mode is the massless one. This means that this is the unique mode living inside the membrane.

\subsection{The Transfer Matrix}

Here we must give the details of the program used to compute the transmission coefficients by transfer matrix. First of all, we consider as an example the simple case given by the double well barrier as shown in the Fig.\ref{fig:multistep1}.

\FIGURE{
\centerline{\psfig{figure=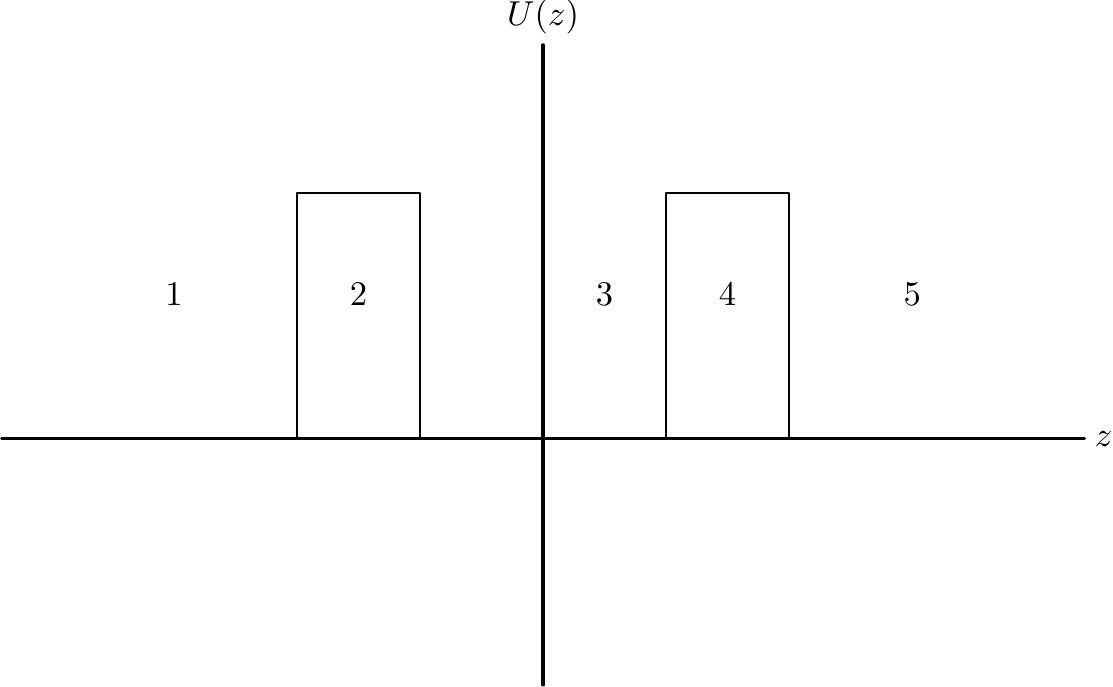,angle=0,height=4cm}}
\caption{The double potential barrier.}\label{fig:multistep1}
}
The solution of the Schroedinger equation for each region is given by
$$
\psi_i(z)=A_ie^{ik_iz}+B_ie^{-ik_iz},\quad k_i=\sqrt{2(E-U_i)},\quad i=1,2,3,4,5.
$$
and is well known from quantum mechanics that it exhibit the following resonance profile as shown in Fig.\ref{fig:square}:

\FIGURE{
\centerline{\psfig{figure=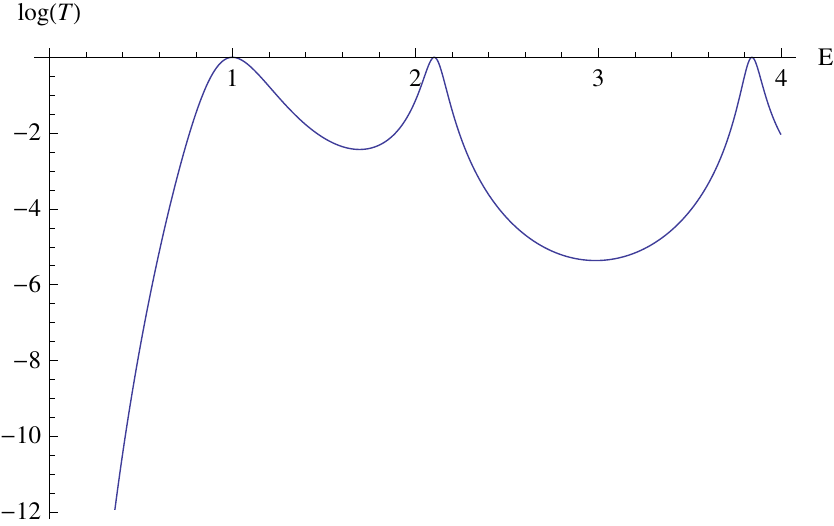,angle=0,height=4cm}}
\caption{Resonance in the double potential barrier.}\label{fig:square}
}
The idea therefore is to apply the results above to a more general case. We are interested in apply that idea for an equation like
\begin{equation}
\{-\frac{d^2}{dz^2}+U(z)\}\psi=\lambda\psi,
\end{equation}
with potential given by the Fig.\ref{fig:multistep2}.

\FIGURE{
\centerline{\psfig{figure=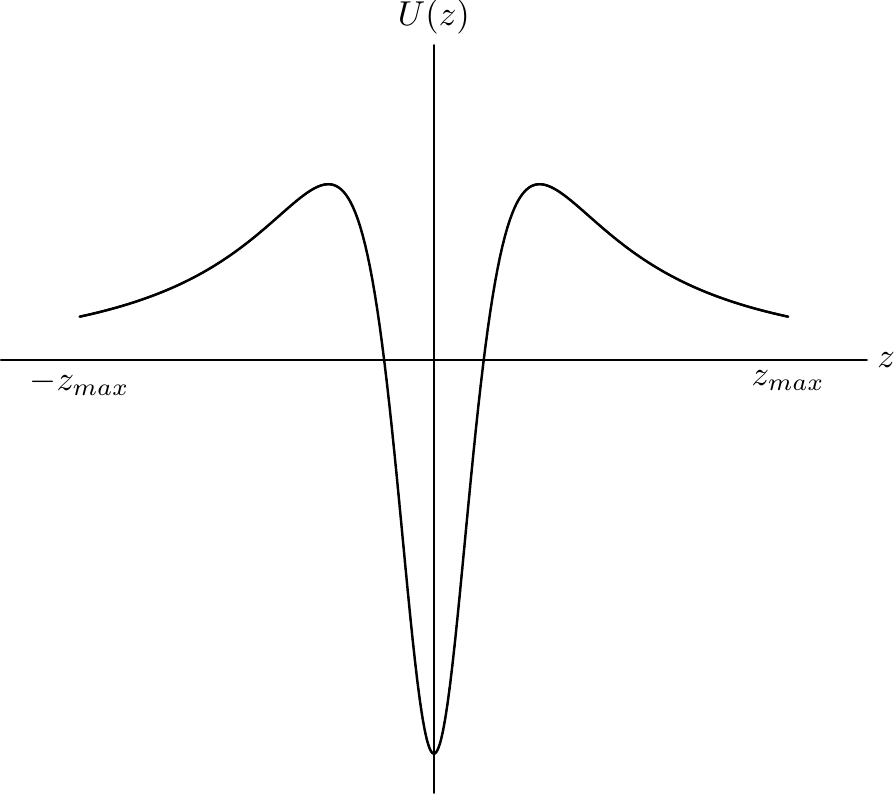,angle=0,height=5cm}}
\caption{General potential with parity symmetry with $\lim_{z \rightarrow \pm\infty}U(z)=0$. }\label{fig:multistep2}
}

In order to solve this, some authors have tried to find numerically relative probabilities defined as follows:
\begin{equation}
 P(m)=\frac{\int_{-z_b}^{z_b} |\psi(z)|^2 dz}
                 {\int_{-z_{max}}^{z_{max}} |\psi(z)|^2 dz}.
 \label{Probability}
\end{equation}
As said before, the opinion of the present authors is that the above prescription hides the physics of the problem and the right approach should use plane waves, looking for transmission
 coefficients.  For this purpose  the potential can be approximated by a series of barriers showed in the Fig.\ref{fig:multistep3}:

\FIGURE{
\centerline{\psfig{figure=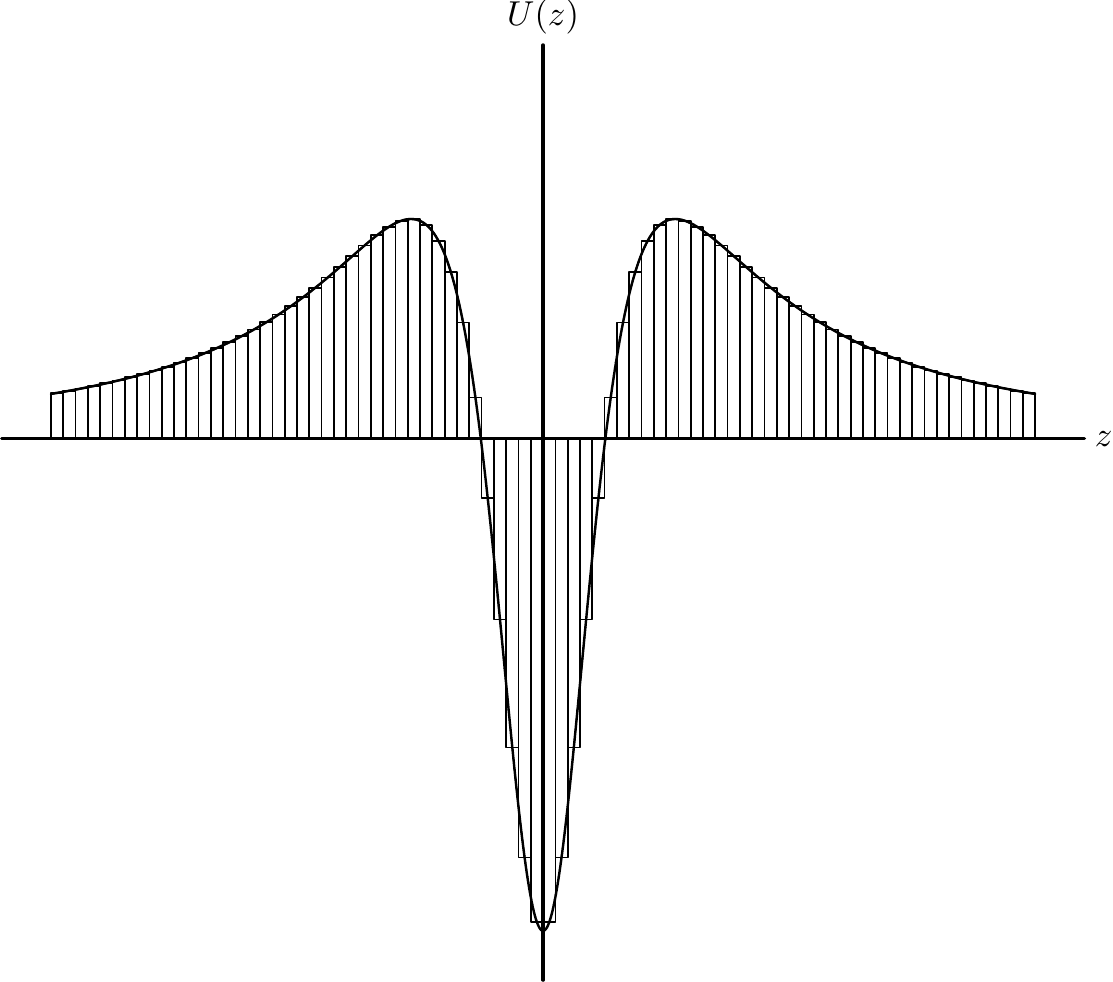,angle=0,height=6cm}}
\caption{Potential approximation by barrier.}\label{fig:multistep3}
}
The Schroedinger equation must be solved for each interval $z_{i-1}<z<z_i$, where we have approximate the potential by 
\bq
U(z)=U(\overline{z}_{i-1})=U_{i-1},\quad\overline{z}_{i-1}=(z_i+z_{i-1})/2.
\ee
with graphic given by the Fig.\ref{fig:multistep4}

\FIGURE{
\centerline{\psfig{figure=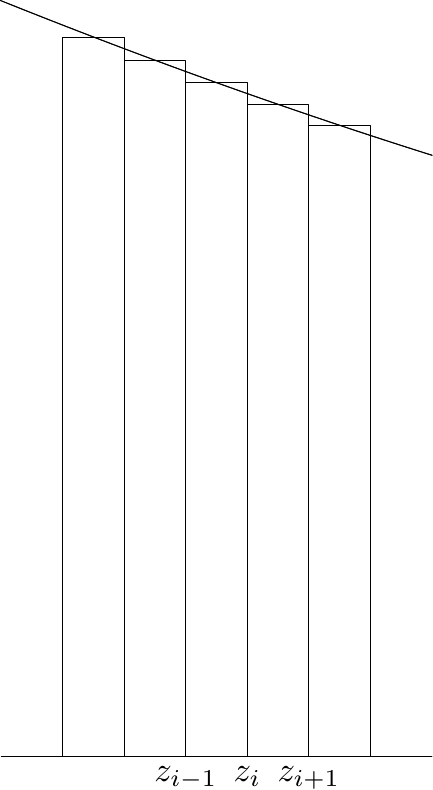,angle=0,height=6cm}}
\caption{The multistep regions.}\label{fig:multistep4}
}
The solution in this interval is
\bq
\psi_{i-1}(z)=A_{i-1}e^{ik_{i-1}z}+B_{i-1}e^{-ik_{i-1}z},\quad k_{i-i}=\sqrt{\lambda-U_{i-1}},
\ee
and the continuity of the $\psi_{i-1}(z)$ and $\psi'_{i-1}(z)$ at $z=z_i$ give us
\bq
\left(
\begin{array}{c}
 A_i\\ 
B_i
\end{array} \right)
=
M_i\left(
\begin{array}{c}
 A_{i-1}\\ 
B_{i-1}
\end{array} \right).
\ee
In the above equation we have that
\bq
M_i=
\frac{1}{2k_i}\left(\begin{array}{cc}
 (k_i+k_{i-1})e^{-i(k_i-k_{i-1})z_i}& (k_i-k_{i-1})e^{-i(k_i+k_{i-1})z_i} \\ 
(k_i-k_{i-1})e^{i(k_i+k_{i-1})z_i} & (k_i+k_{i-1})e^{i(k_i-k_{i-1})z_i}
\end{array} \right)
\ee
and performing this procedure iteratively we reach
\bq
\left(
\begin{array}{c}
 A_N\\ 
B_N
\end{array} \right)
=
M\left(
\begin{array}{c}
 A_{0}\\ 
B_{0}
\end{array} \right),
\ee
where,
\bq
M=M_NM_{N-1}\cdots M_{2}M_1.
\ee
The transmission coefficient is therefore given by
\bq
T=1/|M_{22}|^2
\ee
and this expression is what we must to compute as a function of $\lambda$, which in our case is $m^2$. In order to obtain the numerical resonance values we choose the $z_{max}$ to satisfy $U(z_{max})\sim 10^{-4}$ and $m^2$ runs from $U_{min}=U(z_{max})$ to $U_{max}$ (the maximum potential value). We divide $2z_{max}$ by $10^4$ or $10^5$ such that we have $10^4+1$ or $10^5+1$ transfer matrices.

\section{The Gravity Resonances}

As said in the introduction we are not going to study the localization of zero modes of fields. This have been done extensively in the literature. Here we must concentrate in the possibility of the existence of resonant modes. We consider the two backgrounds found in the last section.

In order to compute resonant massive modes we just need the Schroedinger like equation that is obtained for each kind of field. First of all we must consider resonances of the gravity field with the dilaton contribution. 
We must consider the fluctuation  
\begin{equation}
G_{MN}'=G_{MN}+h_{MN},
\end{equation}
where $h_{MN}$ represents the graviton in the axial gauge $h_{5M}=0$. We must also assume that 
\begin{equation}
h_{\mu}^{\mu}=\partial_{\mu}h^{\mu\nu}=0,
\end{equation}
because we are interested on the transverse modes. With these considerations we can obtain the equation,
\begin{equation}
\left\{-e^{2(A-B)}\frac{\partial^2}{\partial y^2}+e^{2(A-B)}B' \frac{d}{dy}+2e^{2(A-B)}\left(A^{''}-A'B'+2(A')^2\right)-\partial^2\right\}h_{\mu\nu}=0,
\end{equation}
with the definition $\partial^2\equiv \eta^{\mu\nu}\partial_{\mu}\partial_{\nu}$. Performing now the separation of variables 
$h_{\mu\nu}= \overline{h}_{\mu\nu}(x)\psi(y)$, we get the equation
\begin{equation}
-\psi^{''}(y)+B'\psi'(y)+2\left(A^{''}-A'B'+2(A')^2\right)\psi(y)=m^2e^{2(B-A)}\psi(y).
\end{equation}
It has been shown in \cite{Kehagias:2000au} that the zero mode is localized. Using now the relation between $A$ and $B$ and our previous transformations \equ{potential} we get the Schroedinger equation
\begin{equation}
\{-\frac{d}{dz^2}+\overline{U}\}\overline{\psi} =m^2\overline{\psi},
\end{equation}
with potential given by
\begin{equation}
\overline{U}=\frac{3}{2}e^{3A/2}\left(A^{''}+\frac{9}{4}(A')^2\right).
\end{equation}

 Now we study resonances in a dilaton free background. The strategy here is the same as that used to find the background solution without the dilaton contribution. The final result is obtained just by setting $B=0$ to arrive at the equation 
\begin{equation}
\left\{-e^{2A(y)}\frac{\partial^2}{\partial y^2}+2e^{2A(y)}\left(A^{''}+2(A')^2\right)-\partial^2\right\}h_{\mu\nu}=0,
\end{equation}
with the same definitions as before. We get therefore the equation
\begin{equation}
\left\{-\frac{d^2}{dy^2}+\left(2A^{''}+4(A')^2\right)\right\}\psi(y)
={m^2}e^{-2A(y)}\psi(y).
\end{equation}
In order to study the massive spectrum and resonances we must transform the above equation into a Schroedinger equation. This is easily obtained with the transformations \equ{potential} and we obtain
\begin{equation}
\left\{-\frac{d^2}{dz^2}+\overline{U}(z)\right\}\overline{\psi}(z)=m^2\overline{\psi}(z),
\end{equation}
with potential $\overline{U}(z)$ given by
\begin{equation}
\overline{U}(z)=\frac{3}{4}e^{2A}\left(2A^{''}+5(A')^2\right).
\end{equation}
We show in the Fig.\ref{fig:grav}  the graphics of the gravity potential. As said in the second section, this potential has the general behaviour $\lim _{z \rightarrow \pm\infty}U(z)=0$. We must compute transmission coefficients to study the possibility of massive gravity with the method of transfer matrix. We show in Fig.\ref{fig:grav-logt} the result graphically. 

\FIGURE{
\centerline{\psfig{figure=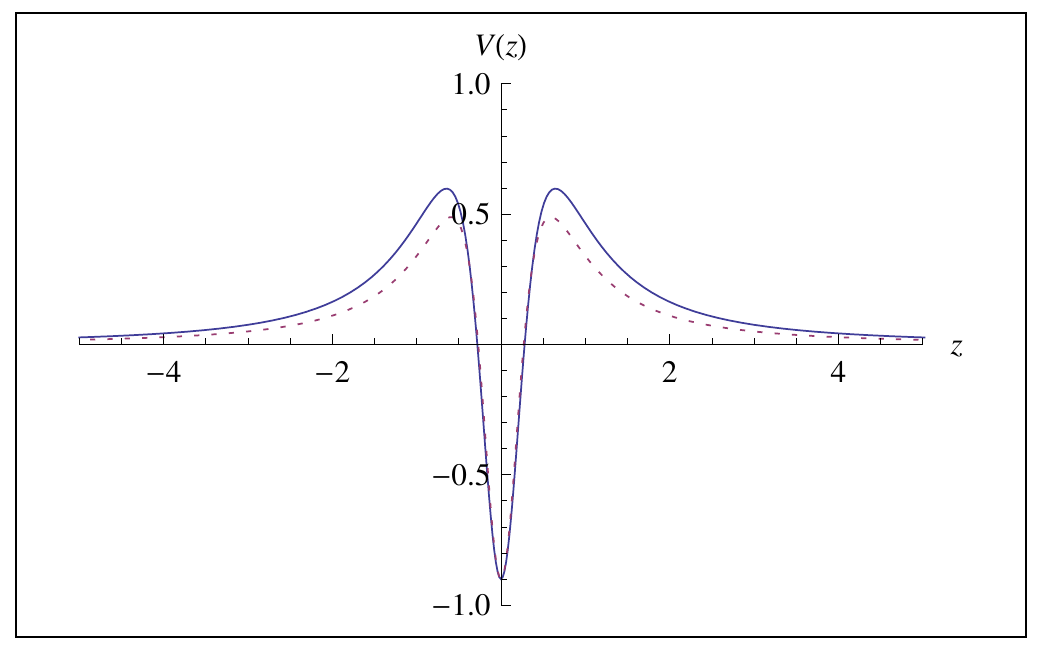,angle=0,height=4cm}}
\caption{Potential of the Schr\"odinger like equation of gravitational field with dilaton (lined) and without dilaton (dotted).\label{fig:grav}}
}

\FIGURE{
\centerline{\psfig{figure=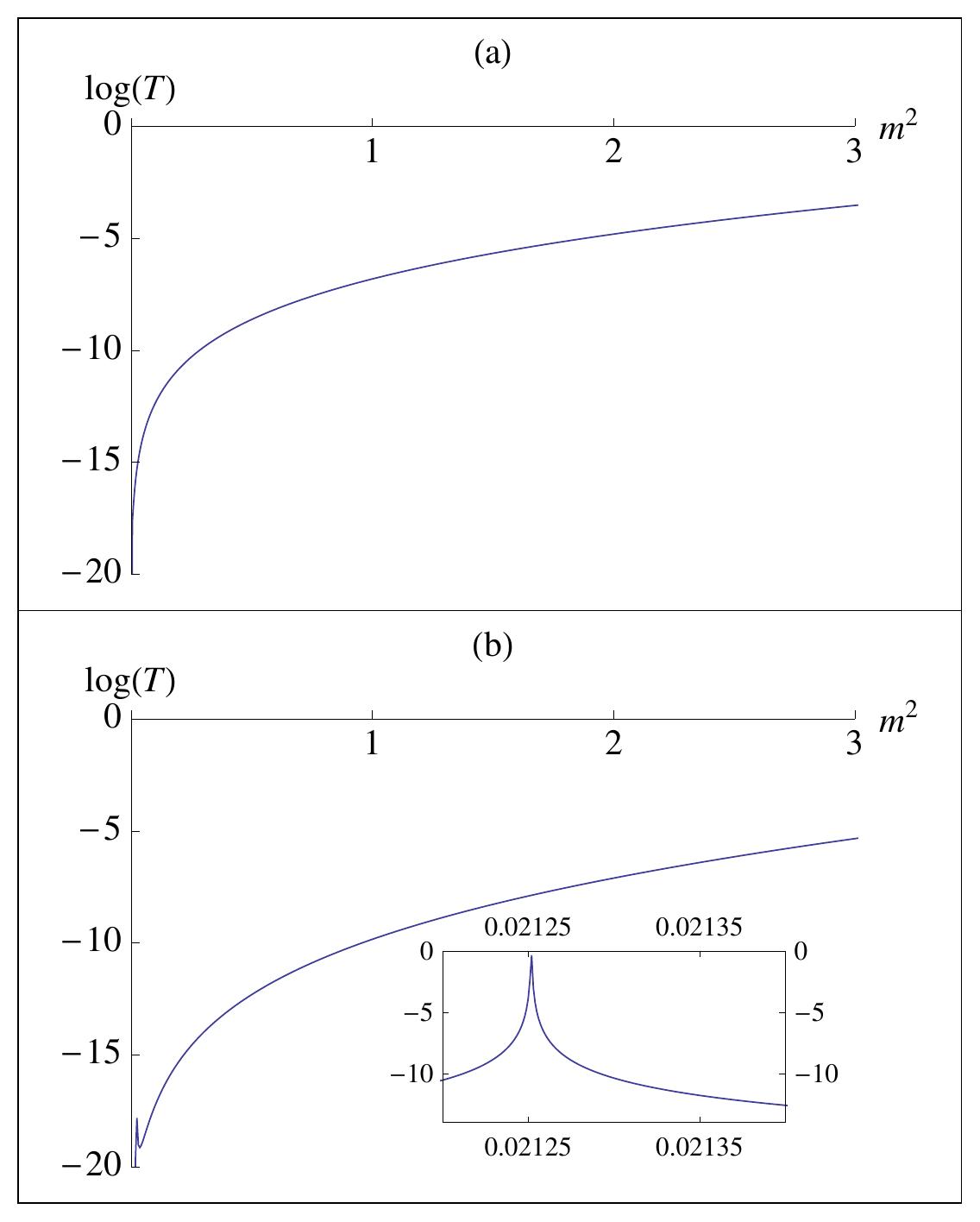,angle=0,height=8cm}}
\caption{Logarithm of the transmission coefficient (a) without dilaton  and (b)with dilaton.\label{fig:grav-logt}}
}

From the Fig.\ref{fig:grav-logt}(b) we can see a peak of probability for massive gravitons near $m^2=0.02125$. Therefore this massive mode is to be expected to interact with the membrane. We should point that we have the absence of resonances without dilaton, as we can see in Fig.\ref{fig:grav-logt}(a). 
 This result agree with \cite{Gremm:1999pj} although the author has used a slightly different topological defect. It is important to note that the appearance of resonances is model dependent.

\section{The Scalar Field Resonances}

In this section we study the massive modes of scalar fields. This can be considered the simplest application of the method.

Initially we take the action for the scalar field $\Phi$ coupled to
gravity,
\begin{equation}\label{acao}
\frac{1}{2}\int d^4xdy \sqrt{-G}e^{-\lambda\pi}G^{MN}\partial_M\Phi\partial_N\Phi,
\end{equation}
where the indexes $M,N$ go from $0$ to $4$. We continue by finding
the equations of motion,
\begin{equation}\label{eqmov}
\eta^{\mu\nu}\partial_\mu\partial_\nu\Phi+e^{-2A-B+\lambda\pi}\partial_y[e^{4A-B-\lambda\pi}\partial_y\Phi]=0,
\end{equation}
where we have separated the extra dimension from the
others. In the next step we use the following separation of
variables:
\begin{equation}\label{eq}
\Phi(x,y)=\chi(x)\psi(y).
\end{equation}
Writing $\Phi$  as above we arrive at the following equation for the
$y$ dependence:
\begin{equation}
\{\frac{d^2}{dy^2}+(4A'-B'-\lambda\pi)\frac{d}{dy}\}\psi=-m^2e^{2(B-A)}\psi.\label{ydep}
\end{equation}
Now we must use our relation between $B$ and $A$ and the transformation \equ{potential} to get our Schroedinger equation
\begin{equation}\label{schrop}
\left\{-\frac{d^2}{dz^2}+\overline{U}(z)\right\}\overline{\psi}=m^2\overline{\psi},
\end{equation}
where the potential $\overline{U}(z)$ assumes the form
\begin{eqnarray}
 &&\overline{U}(z)=e^{3A/2}\left((\frac{\alpha^2}{4}-\frac{9}{64})A'(y)^2-(\frac{\alpha}{2}+\frac{3}{8})A''(y)\right),  \nonumber 
\end{eqnarray}
where $\alpha=-15/4-\lambda\sqrt{3M^3}$.

For the dilaton free case we just fix $B=\pi=0$ to obtain the equation o motion
\begin{equation}\label{eqmov1}
\eta^{\mu\nu}\partial_\mu\partial_\nu\Phi+e^{-2A(y)}\partial_y[e^{4A(y)}\partial_y\Phi]=0,
\end{equation}
and performing again the separation $\Phi(x,y)=\chi(x)\psi(y)$ we get for the $y$ dependence:
\begin{equation}
\{\frac{d^2}{dy^2}+4A'\frac{d}{dy}\}\psi=-m^2e^{-2A}\psi.\label{ydep1}
\end{equation}
Now using \equ{potential} we obtain the Schroedinger equation 
\begin{equation}\label{schrop1}
\left\{-\frac{d^2}{dz^2}+\overline{U}(z)\right\}\overline{\psi}=m^2\overline{\psi},
\end{equation}
where the potential $\overline{U}(z)$ assumes the form
\begin{equation}
\overline{U}(z)=e^{2A}\left[\frac{15}{4}(A')^2+\frac{3}{2}A''\right].
\end{equation}
We show in Figs.\ref{fig:scalar} and \ref{fig:scalar-logt} the graphics of the potential and the logarithm of the transmission coefficient for the scalar field. As pointed in the last
section the presence of the dilaton can induce the appearance of resonant peaks. In Fig.\ref{fig:scalar-logt}(c) we see a very rich structure of resonances for $\lambda\sqrt{3M^3}=40$. 
This indicates that, for this model, some massive scalar field have a peak of probability to interact with the membrane.

\FIGURE{
\centerline{\psfig{figure=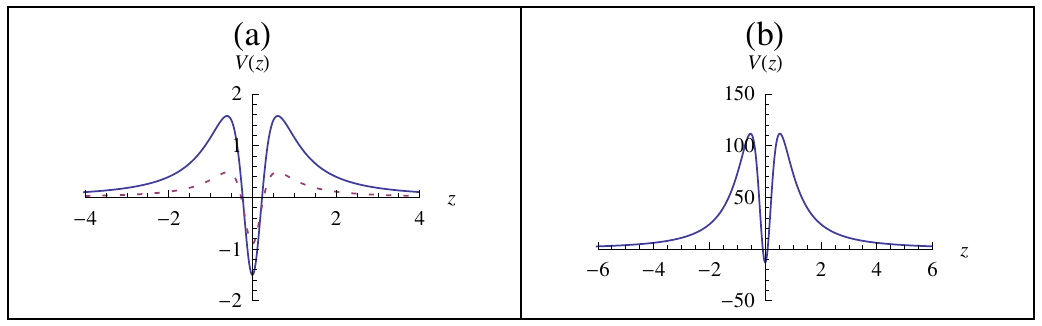,angle=0,height=4cm}}
\caption{Potential of the Schr\"odinger like equation of scalar field scaled by 1/10. (a) with dilaton (lined) for $\lambda\sqrt{3M^3}=2$ and without dilaton (dotted). (b)  with dilaton  for $\lambda\sqrt{3M^3}=40$.\label{fig:scalar}}
}

\FIGURE{
\centerline{\psfig{figure=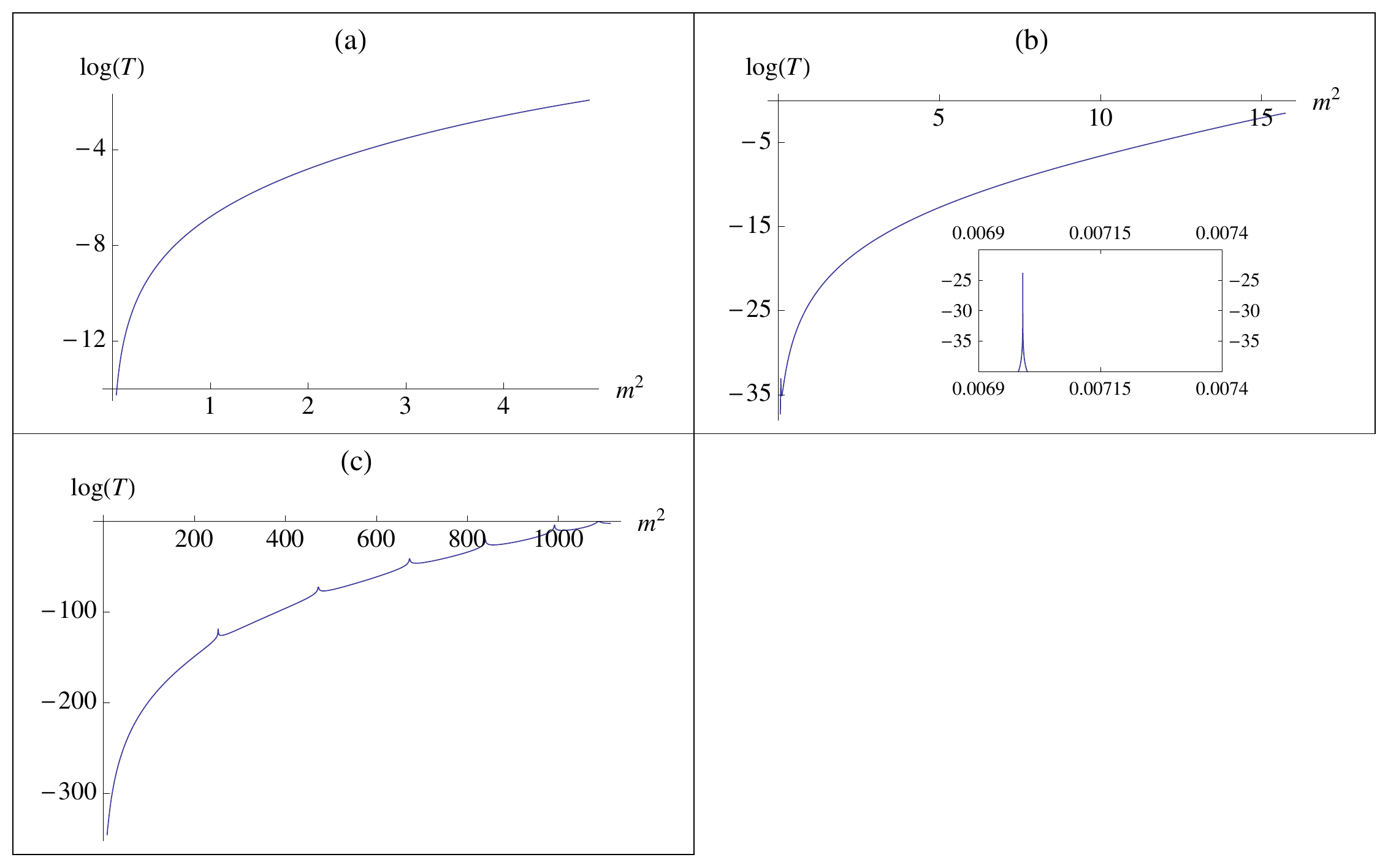,angle=0,height=8cm}}
\caption{Logarithm of the transmission coefficient (a) without dilaton, (b) with dilaton for $\lambda\sqrt{3M^3}=2$  and (c) with dilaton for $\lambda\sqrt{3M^3}=40$.\label{fig:scalar-logt}}
}

\section{The Vector Field Resonances}

Here we must consider the next important field, which is the gauge field. It has been show that this field is localized only in the Dilatonic background. Anyway, for completeness, we give an analyses of the resonances for the dilaton free case. These cases have already been studied in \cite{Kehagias:2000au}\cite{Cruz:2010zz}, but we analyses it here in the light of the Transfer Matrix Method.

The coupling between the dilaton and the gauge field is inspired in string theory and we have for the action
\begin{equation}
S_X=\int d^{D}x\sqrt{-G}e^{-\lambda \pi }[Y_{M_{1}M_{2}}Y^{M_{1}M_{2}}],
\end{equation}
where $Y_{M_{1}M_{2}}=\partial _{[M_{1}}X_{M_{2}]}$ is the field
strength for the $1$-form $X$. The equation of motion is:
\begin{equation}
\partial_{M}(\sqrt{-g}g^{MP}g^{NQ}e^{-\lambda\pi}H_{PQ})=0.
\end{equation}
We can use gauge freedom to fix $X_{y}=\partial^{\mu}X_{\mu}=0$ and we are left with the following two type of terms
\begin{equation}
Y_{\mu_{1}\mu_{2}}=\partial _{[ \mu_{1}}X_{\mu_{2}]},
\end{equation}
\begin{equation}
Y_{y\mu}=\partial _{[ y}X_{\mu}] =\partial_y X_{\mu}=X'_{\mu}.
\end{equation}
The equation of motion takes the form
\begin{eqnarray}
\partial _{\mu_1 }Y^{\mu_1\mu_{2}}
+e^{((-B+\lambda\pi)}\partial _{y}[e^{(2A-B-\lambda\pi)}{X'}^{\mu_2}]=0,
\end{eqnarray}
and performing the same separation of variable we get
\begin{equation}
\{\frac{d^2}{dy^2}-\left( -2{A}'+{B}'+\lambda\pi'\right)\frac{d}{dy}\}\psi\left(
y\right) =-m^{2}e^{2\left( B-A\right) }\psi\left( y\right) . \label{Udilaton}
\end{equation}
Now we can get our Schroedinger equation by performing the transformation \equ{potential} and use the relations between $B,\pi$ and $A$. We arrive at
\begin{eqnarray}
 &&\overline{U}(z)=e^{3A/2}\left((\frac{\alpha^2}{4}-\frac{9}{64})A'(y)^2-(\frac{\alpha}{2}+\frac{3}{8})A''(y)\right),  \nonumber ,
\end{eqnarray}
where $\alpha=-7/4-\lambda\sqrt{3M^3}$

Now we repeat the same lines used throughout all the paper to reach the Dilaton Free case. First of all the action is given by
\begin{equation}
S_X=\int d^{5}x\sqrt{-G}[Y_{M_{1}M_{2}}Y^{M_{1}M_{2}}],
\end{equation}
with the same definitions as before. The equation of motion can be obtained from the previous case by fixing $B=\pi=0$ and we get
\begin{eqnarray}
\partial _{\mu_1 }Y^{\mu_1\mu_{2}}
+\partial _{y}[e^{2A}{X'}^{\mu_2}]=0,
\end{eqnarray}
and performing the same separation of variable we get
\begin{equation}
\{\frac{d^2}{dy^2}+2{A}'\frac{d}{dy}\}\psi\left(
y\right) =-m^{2}e^{-2A }\psi\left( y\right) . \label{Udilaton1}
\end{equation}
Now we can get the Schroedinger equation by performing the transformation \equ{potential} to arrive at
\begin{equation}
\overline{U}=e^{2A}\{\frac{3}{4}A'^2+\frac{1}{2}A''\}.
\end{equation}
We show in the Fig.\ref{fig:vector} graphics of the vector field potential. In the Fig.\ref{fig:logt-vector}  we present the logarithm of the transmission coefficient. 
As commented in the introduction, the coupling with the dilaton induces a rich structure of resonances. Without the dilaton we see that we do not have any resonances. With the inclusion the 
dilaton we see in in Fig.\ref{fig:logt-vector}(b) that we find one resonance peak very close to zero. When we enlarge the value of the coupling constant we get six peaks of resonances, a result which corroborate our claim that very massive modes can also interact with the membrane.

\FIGURE{
\centering
\centerline{\psfig{figure=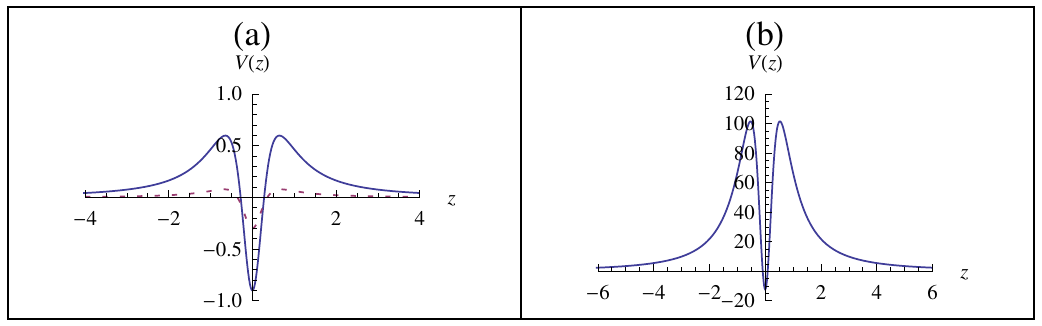,angle=0,height=4cm}}
\caption{Potential of the Schroedinger like equation of vector field scaled by 1/10. (a) With dilaton (lined) for $\lambda\sqrt{3M^3}=2$ and without dilaton (dotted), (b) with dilaton  for $\lambda\sqrt{3M^3}=40$\label{fig:vector}}
}

\FIGURE{
\centering
\centerline{\psfig{figure=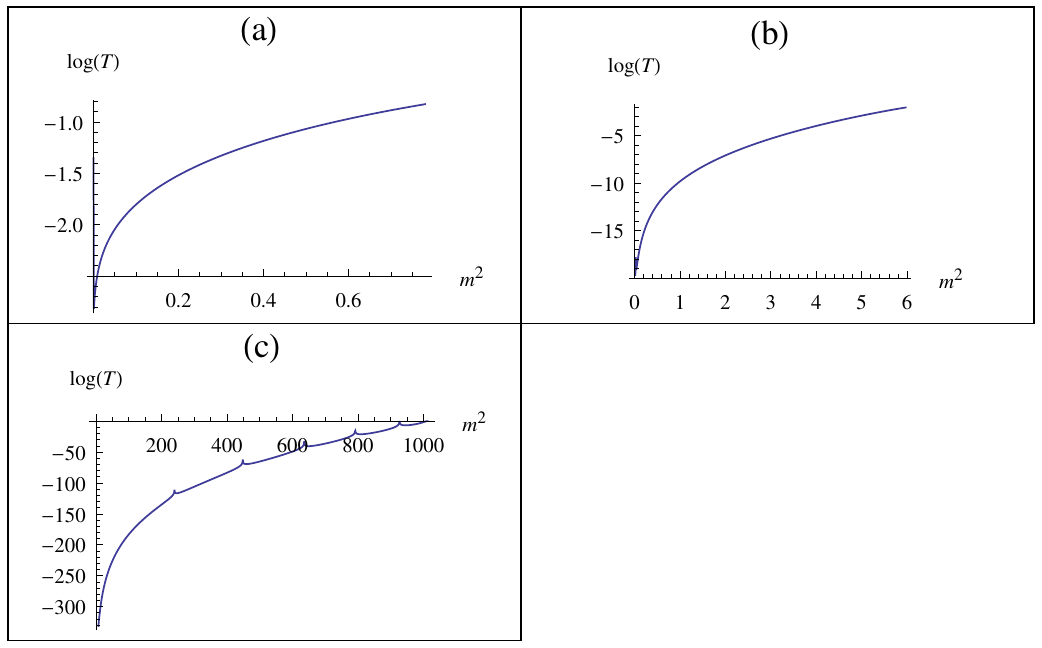,angle=0,height=8cm}}
\caption{Logarithm of the transmission coefficient. (a) Without dilaton, (b) with dilaton for $\lambda\sqrt{3M^3}=2$  and (c) with dilaton for $\lambda\sqrt{3M^3}=40$\label{fig:logt-vector}}
}

\section{The Kalb-Ramond Resonances}

The dilaton coupling here is the same as that for the scalar and vector fields. The reason is that the dilaton can be seen as the radius of a compactified extra dimension and all the fields get an exponential factor coming from the measure. Therefore we have for the action
\begin{equation}
S_X=\int d^{5}x\sqrt{-G}e^{-\lambda \pi }Y_{M_{1}M_{2}M_{3}}Y^{M_{1}M_{2}M_{3}},
\end{equation}
where $Y_{M_{1}M_{2}M_{3}}=\partial _{[M_{1}}X_{M_{2}M_{3}]}$ is the field strength for the $2$-form $X$. The new equation of motion is:
\begin{equation}
\partial_{M}(\sqrt{-g}g^{MP}g^{NQ}g^{LR}e^{-\lambda\pi}H_{PQR})=0.
\end{equation}
We can use gauge freedom to fix $X_{\mu_{1}y}=\partial^{\mu_{1} }X_{\mu{_{1}}\mu_{2}}=0$ and we are left with the following two type of terms
\begin{equation}
Y_{\mu_{1}\mu_{2}\mu_{3}}=\partial _{[ \mu_{1}}X_{\mu_{2}\mu_{3}]},
\end{equation}
\begin{equation}
Y_{y\mu_{1}\mu_{2}}=\partial _{[ y}X_{\mu_{1}\mu_{2}}] =\partial_y X_{\mu_{1}\mu_{2}}=X'_{\mu_{1}}{\mu_{2}}.
\end{equation}
The equation of motion takes the form
\begin{eqnarray}
\partial _{\mu_{1} }Y^{\mu_{1} \mu_{2}\mu_{3}}
+e^{(2A-B+\lambda\pi)}\partial _{y}[e^{-(B+\lambda\pi)}{X'}^{\mu_{2}\mu_{3}}]=0,
\end{eqnarray}
and we can now separate the $y$ dependence of the field using
\begin{equation}
X^{\mu_{1}\mu_{2}}\left( x^{\alpha },y\right) =B^{\mu_{1}\mu_{2}}\left(
x^{\alpha }\right) \psi\left( y\right).
\end{equation}
Defining $Y^{\mu_{1}\mu_{2}}=\tilde{Y}^{\mu_{1}\mu_{2} }\psi$, where $\tilde{Y}$ stands for the four dimensional field strength, we get
\begin{equation}
 \partial _{\mu_{1} }\tilde{Y}^{\mu_{1} \mu_{2}}+m^2B^{\mu_{2}\mu_{1}}=0
\end{equation}
and the differential equation which give us information about the extra dimension, namely
\begin{equation}\label{zero}
\frac{d^{2}\psi(y)}{dy^{2}}-(\lambda\pi^{\prime}(y)+B^{\prime}(y))\frac{d\psi(y)}{dy}=-m^{2}e^{2(B(y)-A(y))}\psi(y).
\end{equation}
Using our previous relation between $B$,$\pi$ and $A$ and performing the transformation \equ{potential} we get
\begin{equation}\label{schro}
\left\{-\frac{d^2}{dz^2}+\overline{U}(z)\right\}\overline{\psi}=m^2\overline{\psi},
\end{equation}
where the potential $\overline{U}(z)$  assumes the form,
\begin{equation}\label{pot_reson}
\overline{U}(z)=e^{\frac{3}{2}A}\left[\left(\frac{\alpha^2}{4}-\frac{9}{64}\right)(A')^2-\left(\frac{\alpha}{2}+\frac{3}{8}\right)A''\right].
\end{equation}
where,
\begin{equation}
\beta=\frac{\alpha}{2}+\frac{3}{8},\,\alpha=\frac{1}{4}-\sqrt{3M^3}\lambda.
\end{equation}

Now we introduce the action for the Kalb-Ramond field without the dilaton as
\begin{equation}
S_X=\int d^{5}x \sqrt{-G}[H_{MNL}H^{MNL}],
\end{equation}
and we can again obtain the equations of motion setting $B=\pi=\lambda=0$. The
result is given by:
\begin{equation}
\partial_{\mu_{1}} Y^{\mu_{1} \mu_{2} \mu_{3}}\psi(y)-e^{2A}\frac{d^2\psi(y)}{dy^2}X^{\mu_{2} \mu_{3}}=0.
\end{equation}
The function $\psi(y)$ carry all information about the extra dimension and obeys the following equation:
\begin{equation}\label{motion}
\frac{d^2\psi(y)}{dy^2}=-m^2e^{-2A(y)}\psi(y).
\end{equation}
We must make now the standard transformation \equ{potential} to get
\begin{equation}\label{sch}
\left\{-\frac{d^2}{dz^2}+\overline{U}(z)\right\}\overline{\psi}(z)=m^2\overline{\psi},
\end{equation}
where
\begin{equation}
\overline{U}(z)=e^{2A}[\frac{1}{4}(A')^2+\frac{1}{2}(A'')].
\end{equation}

We show in the Fig.\ref{fig:tensor} the graphics of the Kalb-Ramond field potential. In Fig.\ref{fig:logt-tensor} we show the logarithm of the transmission coefficient. 
As in the scalar and vector field, the resonance only exist for large coupling constant. Note that, very similar to the vector field case, the number of resonances increase very much when consider a large value for the coupling constant.

\FIGURE{
\centering
\centerline{\psfig{figure=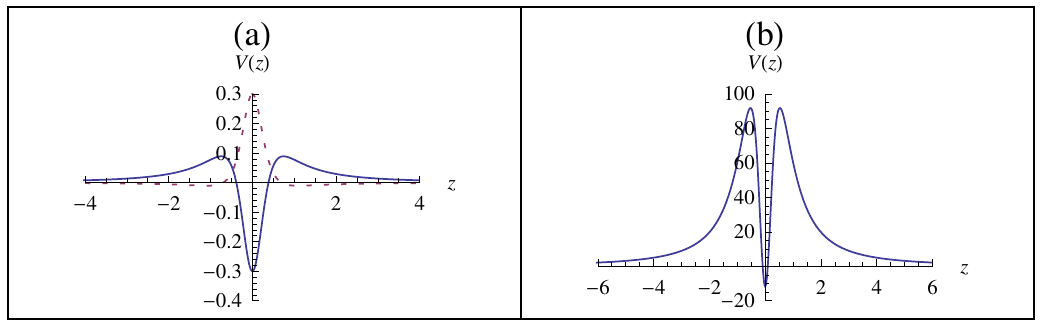,angle=0,height=4cm}}
\caption{Potential of the Schr\"odinger like equation of Kalb-Ramond field scaled by 1/10. (a) With dilaton (lined) for $\lambda\sqrt{3M^3}=2$ and without dilaton (dotted), (b) with dilaton  for $\lambda\sqrt{3M^3}=40$.\label{fig:tensor} }
}

\FIGURE{
\centering
\centerline{\psfig{figure=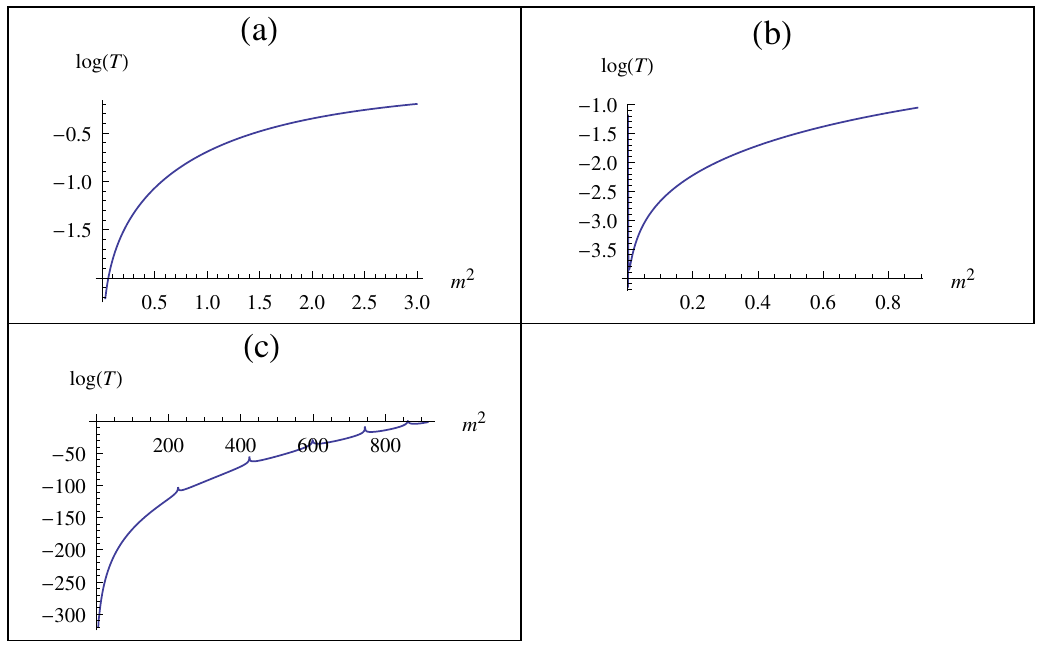,angle=0,height=8cm}}
\caption{Logarithm of the transmission coefficient. (a) Without dilaton, (b) with dilaton for $\lambda\sqrt{3M^3}=2$  and (c) with dilaton for $\lambda\sqrt{3M^3}=40$.\label{fig:logt-tensor} }
}

\section{The Fermion Resonances}

The next important case to be studied is the fermion field. The issue of zero mode localization of this field has attracted much attention 
recently \cite{Liu:2009ve,Zhao:2009ja,Liang:2009zzf,Zhao:2010mk,Zhao:2011hg,Li:2010dy,
Castro:2010au,Correa:2010zg,Castro:2010uj,Chumbes:2010xg,Castro:2011pp}. Here we are interested only in the study of resonant modes. They depend 
strongly of the form of the potential and, therefore, of the model considered. Here we must study the model similar to that found  in \cite{Kehagias:2000au} and \cite{Li:2010dy}, 
therefore we must give only a short review on how to arrive in the Schroedinger equation that determines the massive modes. The action considered is slightly modified to include the
dilaton coupling, as in our previous sections. The action is given bellow
\begin{eqnarray}
 S_{1/2} &=& \int d^{5}x\sqrt{-g}e^{-\lambda\pi}\left[\overline{\Psi}\Gamma^{M}D_{M}\Psi
    -\eta\overline{\Psi}F(\phi)\Psi\right],
    \label{fermion field action}
\end{eqnarray}
where $D_M=\partial_M+\omega_M$, $\omega_M$ being the spin connection. The $\Gamma^M=e_{N}^{~~M}\gamma^N$ are the Dirac matrices in the five dimensional curved space-time 
and $e_{N}^{~M}$ are the vielbeins: $e_{A}^{~~M}e_{B}^{~~N}\eta^{AB}=g^{MN}$. After some manipulations with the spin connection we get the equations of motion

\begin{eqnarray}
\left[\gamma^{\mu}\partial_{\mu}+e^{A-B}\gamma^{5}\left(\partial_{y}+2\partial_{y}A\right)-\eta
e^{A}F(\phi)\right]\Psi=0\,,
 \label{Dirac}
\end{eqnarray}
where $\gamma^{\mu}\partial_{\mu}$ is the four-dimensional Dirac
operator on the brane. It is interesting to note here that the dilaton coupling just modifies the equation of motion throughout the $B$ factor of the metric. Therefore the value of $\lambda$ 
will be irrelevant to analyses the resonance spectrum. Now, just as in the gravity case we perform the decomposition
\begin{eqnarray}
\Psi(x,y)=e^{-2A}\left(\sum_{n}\psi_{Ln}(x)f_{Ln}(y)+\sum_{n}\psi_{Rn}(x)f_{Rn}(y)\right).\
\label{fermion decomposition}
\end{eqnarray}

In the above we defined $\gamma^{5}\psi_{Ln}(x)=-\psi_{Ln}(x)$ and
$\gamma^{5}\psi_{Rn}(x)=\psi_{Rn}(x)$ as the left and right-handed fermion fields respectively. Just as pointed in the second section we obtain the equation for a massive field in four dimension 
$$\gamma^{\mu}\partial_{\mu}\psi_{Ln}(x)=m_{n}\psi_{Rn}(x)$$
 and
$$\gamma^{\mu}\partial_{\mu}\psi_{Rn}(x)=m_{n}\psi_{Ln}(x),$$
with mass defined by
\begin{eqnarray}
 \left[\partial_{y} +{\eta} e^{B}F(\phi)\right]
  f_{Ln}(y) &=& \,\,\,\,\,m_{n}e^{B-A}f_{Rn}(y)\,,\label{extradimension1} \\
 \left[\partial_{z}-\eta e^{B}F(\phi)\right]
  f_{Rn}(y) &=& -m_{n}e^{B-A}f_{Ln}(y)\,. \label{extradimension2}
\end{eqnarray}

From the these equations we can obtain the Schroedinger equation for each chirality

\begin{eqnarray}
 \left[-\partial_{z}^{2}+U_{L}(z)\right]f_{L} &=& m^{2}f_{L} \,,
    \\ \label{ScheqLeft}
 \left[-\partial_{z}^{2}+U_{R}(z)\right]f_{R} &=& m^{2}f_{R} \,,
       \label{ScheqRight}
\end{eqnarray}
with potential given by
\begin{eqnarray}
U_{L}(z)=e^{2A-2B}\left(\eta^{2} e^{2B}F^{2}(\phi)-\eta
e^{B}\partial_{y}F(\phi)-\eta
e^{B}(\partial_{y}A)F(\phi)\right)\,,  \label{VzL}  \\
U_{R}(z)=e^{2A-2B}\left(\eta^{2} e^{2B}F^{2}(\phi)+\eta
e^{B}\partial_{y}F(\phi)+\eta e^{B}(\partial_{y}A)F(\phi)\right)\,,
\label{fermioneq}
\end{eqnarray}
with $dz/dy=e^{B-A}$.
With this at hand we can now study resonances. As said in all previous section, to obtain the case without the dilaton we just fix $B=0$. We show in the  Fig.\ref{fig:fermion05},  
Fig.\ref{fig:fermion-logt05},  Fig.\ref{fig:fermion10} and in the Fig.\ref{fig:fermion-logt10}  
the graphics of the  fermion potential and the logarithm of the transmission coefficient for $F(\phi)=\phi$. We note that if the potential looks like the double well barrier, we can observe resonances. In the case of the right fermion it looks like a one single barrier and consequently we cannot observe resonances. In Fig.\ref{fig:fermion-logt05}, with $\eta=0.5$, 
we can see that regardless of the dilaton coupling we do not have any resonances. In Fig.\ref{fig:fermion10}, for $\eta=10$ we see that some peaks appear for the left handed fermion. 
Therefore it looks that the resonant structure here depends mostly on the value of $\eta$.

\FIGURE{
\centerline{\psfig{figure=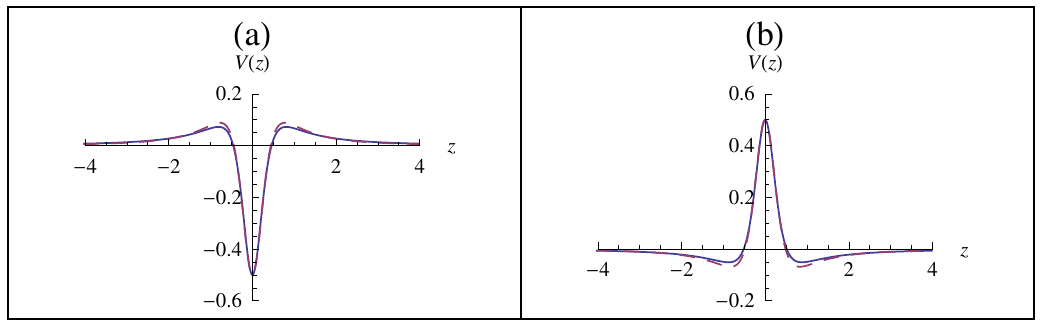,angle=0,height=4cm}}
\caption{Potential of the Schroedinger like equation of fermion with $\eta=0.5$. (a) Left without dilaton (dotted) and with dilaton (lined) and (b) Right without dilaton (dotted) and with 
dilaton (lined).\label{fig:fermion05}}
}

\FIGURE{
\centerline{\psfig{figure=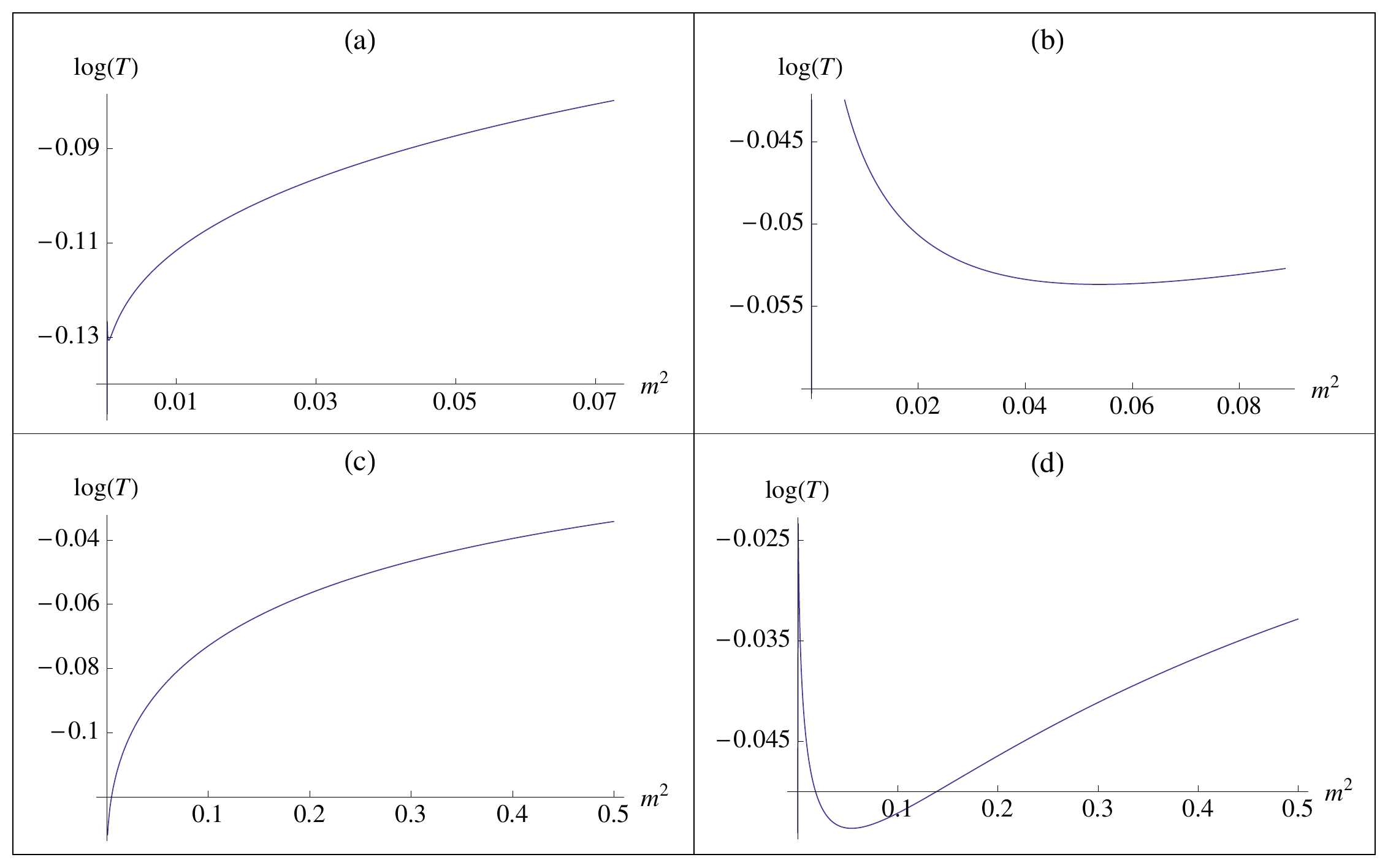,angle=0,height=8cm}}
\caption{Logarithm of the transmission coefficient of fermion with $\eta=0.5$. (a) Left without dilaton, (b) Left with dilaton, (c) Right without dilaton and (d) Right with dilaton.
\label{fig:fermion-logt05}}
}

\FIGURE{
\centerline{\psfig{figure=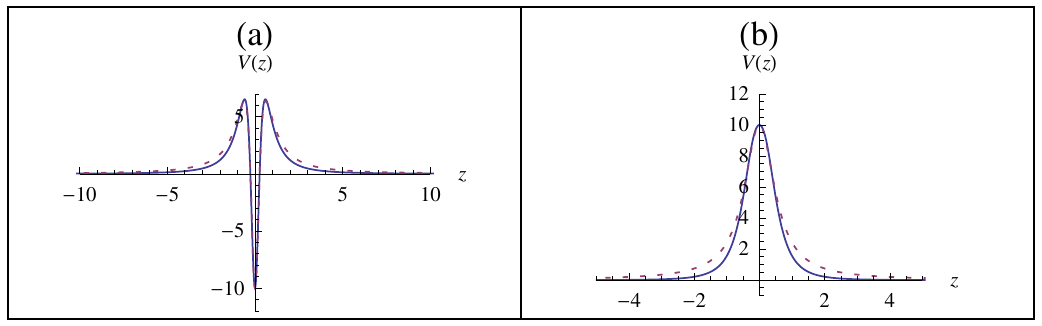,angle=0,height=4cm}}
\caption{Potential of the Schr\"odinger like equation of fermion with $\eta=10$. (a) Left without dilaton (dotted) and with dilaton (lined) and (b) Right without dilaton (dotted) and with 
dilaton (lined).\label{fig:fermion10}}
}

\FIGURE{
\centerline{\psfig{figure=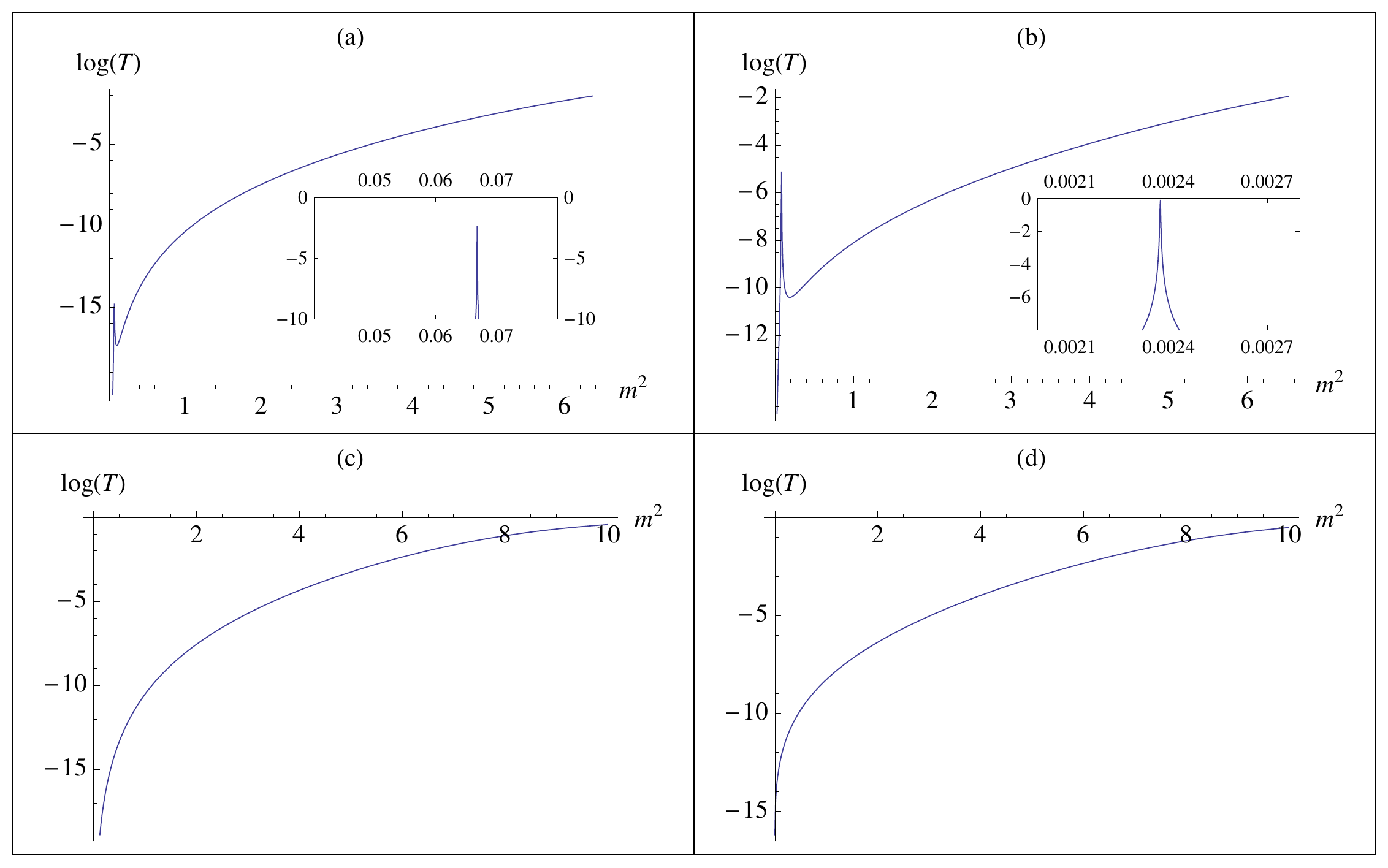,angle=0,height=8cm}}
\caption{Logarithm of the transmission coefficient of fermion with $\eta=10$. (a) Left without dilaton, (b) Left with dilaton, (c) Right without dilaton and (d) Right with dilaton.
\label{fig:fermion-logt10}}
}


\section{Conclusions}

In this work we have addressed the issue of resonances in Randall-Sundrum-like models. This matter has already been studied before for many authors, but the general prescription is to use an non-physical normalization and try to solve the equations of motions. This approach could hide the physics involved in the problem and we propose to study it by the transfer matrix method of computing transmission coefficients. As pointed in the introduction this is a common technique used in condensed matter systems. It is used for example to find the peaks of currents in a given semi-conductor. Roughly speaking, the method consists in taking the potential of the problem and approximate it by a series of potential barriers. Then, for each barrier (for each interval) it is computed interactively the transfer matrix, obtaining finally the total transfer matrix. With this method we have computed resonances of gravity, fermions and gauge fields. 

For the gravity field, we show in the Fig.\ref{fig:grav-logt} the graphic of the logarithm of the transmission coefficient. From that picture we can see that we have no peak of probability for massive gravitons in Fig.\ref{fig:grav-logt}(a). This correspond to dilaton free case.Therefore a massive mode is not expected to interact with the membrane.  For the dilaton case, Fig.\ref{fig:grav-logt}(b) we find a resonance peak. It is important to note that the appearance of resonances is model dependent.  For the simplest case, the scalar field, we show in Fig.\ref{fig:scalar-logt} the graphic of the logarithm of the transmission coefficient. The presence of the dilaton can induce the appearance of resonant peaks. In Fig.\ref{fig:scalar-logt}(c) we see a very rich structure of resonances for $\lambda\sqrt{3M^3}=40$. This indicates that, for this model, some massive scalar field have a peak of probability to interact with the membrane. In the case of the vector gauge field, the coupling with the dilaton induces a rich structure of resonances. Without the dilaton we see that we do not have any resonances. With the inclusion the dilaton we see in in Fig.\ref{fig:logt-vector}(b) that we find one resonance peak very close to zero. When we enlarge the value of the coupling constant we get six peaks of resonances, a result which corroborate our claim that very massive modes can also interact with the membrane. The Kalb-Ramond field behaves as the scalar and vector fields, i.e., the resonance only exist for large coupling constant. We note that, very similar to the vector field case, the number of resonances increase very much when consider a large value for the coupling constant. For the case of fermionic fields we note that if the potential looks like the double well barrier, we can observe resonances. In the case of the right fermion, it looks like a one single barrier and consequently we cannot observe resonances. In Fig.\ref{fig:fermion-logt05}, with $\eta=0.5$, we can see that, regardless of the dilaton coupling, we do not have any resonances. In Fig.\ref{fig:fermion10}, for $\eta=10$ we see that some peaks appear for the left handed fermion. Therefore it looks that the resonant structure here depends mostly on value of $\eta$.

Furthermore, there are some applications of the models studied here related to AdS/CFT correspondence: there are evidences supporting the idea of domain wall/QFT, i.e., a 
correspondence between gauged supergravities and quantum field theories in domain walls. Important applications are related to the study of quark-gluon plasma via gravity-duals. Therefore, we hope to adapt this numerical method to attack such kind of problems.

\section*{Acknowledgment}

We would like to thank the Laborat\'orio de \'Oleos pesados and Goethe-Institut Berlin for the hospitality. We acknowledge Linda Fuchs for help with the manuscript. We also acknowledge the 
financial support provided by Funda\c c\~ao Cearense de Apoio ao Desenvolvimento Cient\'\i fico e Tecnol\'ogico (FUNCAP), the Conselho Nacional de Desenvolvimento Cient\'\i fico e Tecnol\'ogico (CNPq) and FUNCAP/CNPq/PRONEX.

\end{document}